**Brun-Type Formalism for Decoherence in Two Dimensional Quantum Walks**

**CLEMENT AMPADU**

31 Carrolton Road
Boston, Massachusetts, 02132
U.S.A.
e-mail: drampadu@hotmail.com

**Abstract**

We study decoherence in the quantum walk on the xy-plane. We generalize the method of decoherent coin quantum walk, introduced by [T.A. Brun, et.al, Phys.Rev.A 67 (2003) 032304], which could be applicable to all sorts of decoherence in two dimensional quantum walks, irrespective of the unitary transformation governing the walk. As an application we study decoherence in the presence of broken line noise in which the quantum walk is governed by the two-dimensional Hadamard operator.

## 1. Introduction

As is well known the physical implementation of the quantum walk faces many obstacles including environmental noise and imperfections collectively known as *decoherence.* Apart from the review on the decoherent quantum walk given by the authors in reference [1], in [2] the authors study the discrete-time quantum walk on the $N-$cycle subject to decoherence both on the coin and the position degree of freedom, for the bipartite system considered by them, their analysis show that exposure to any nonzero level of persistent decoherence causes the system to behave asymptotically like a purely classical system. In [3] the authors investigate the impact of decoherence and static disorder on the dynamics of quantum particles moving in a periodic lattice, in particular for slowly changing lattice parameters, a fast ballistic spread was observed, in the case of dynamical disorder, a diffusive spread was observed, and in the case of lattices with static disorder, Anderson localization was observed.

The global chirality distribution of the quantum walk on the line when decoherence is introduced either through simultaneous measurements of the chirality and particle position, or as a result of broken links is studied in [4]. The first mechanism drives the system towards a classical diffusive behavior. This is used to build new quantum games, similar to the so-called spin-flip game. The second mechanism involves two different possibilities: (a) All the quantum walk links have the same probability of being broken. (b) Only the quantum walk links on a half-line are affected by random breakage. In case (a) the decoherence drives the system to a classical Markov process, whose master equation is equivalent to the dynamical equation of the quantum density matrix. This is not the case in (b) where the asymptotic global chirality distribution unexpectedly maintains some dependence with the initial condition. Explicit analytical equations are obtained for all cases. The classicalization of a decoherent discrete-time quantum walk on a line or an n-cycle is demonstrated in various ways that do not necessarily provide a geometry-independent description in [5]. It is shown that the degree of quantum correlations between the coin and position degrees of freedom, quantified by a measure based on the disturbance induced by local measurements provides a suitable measure of classicalization across both type of walks. Applying this measure to compare the two walks, the authors find that cyclic quantum walks tend to classicalize faster than quantum walks on a line because of more efficient phase randomization due to the self-interference of the two counter-rotating waves, in addition they model the noise as acting on the coin and given by the squeezed generalized amplitude damping (SGAD) channel, which generalizes the generalized amplitude damping channel. An intrinsically stable, deterministic implementation of discrete quantum walks with single photons in space is presented in [6]. The number of optical elements required scales linearly with the number of steps. The authors measure walks with up to six steps and explore the quantum-to-classical transition by introducing tunable decoherence. In addition they also investigate the effect of absorbing boundaries and show in particular that decoherence significantly affects the probability of absorption. The effects of shifting position decoherence, arising from the tunnelingeffect in the experimental realization of the quantum walk, on the one-dimensional discrete time quantum walk is studied in [7]. The authors show that in the regime of this type of noise the quantum behavior of the walker does not fade, in contrary to the coin decoherence for which the walker undergoes the quantum-to classical transition even for weak noise. Particularly, they show that the quadratic dependency of the variance on the time and also the coin-position entanglement, are preserved in the presence of tunneling decoherence. Furthermore, they present an explicit expression for the probability distribution of decoherent one-dimensional quantum walk in terms of the corresponding coherent probabilities, and show that this type of decoherence smooths the probability distribution. Quantum random walk in a two-dimensional lattice with randomly distributed traps is investigated in [8]. Distributions of quantum walkers are evaluated dynamically for the cases of Hadamard,Fourier, and Grover coins, and quantum to classical transition is examined as a function of the density of the traps. In particular, it is shown that traps act as a serious and additional source of quantum decoherence. In [9] the discrete-time quantum walk on the $N-$cycle is studied, it is shown that there exist a sharp contrast in behavior between quantum walks which are purely coherent and quantum walks which are tainted by even the slightest trace of decoherence. In particular, when exposed to any non-zero level of decoherence on the coin degree of freedom, the authors show that a QW behaves eventually like a classical random walk. In particular, if the decoherence rate $p > 0$, a QW on the N-cycle appears to mix always to a uniform distribution at a rate no faster than a classical random walk. In [10] it is shown that the quantum walk on the circle in phase space involving just one walker can be implemented via circuit quantum electrodynamics with or without open systems effects using realistic parameters and find that the signature of the quantum walk is evident under these conditions. In particular the authors show that direct homodyne measurements over few quadratures reveal an unambiguous quantum walk signature, which implies full tomography may not be required. In particular, it implies that the emergence of the random walk can be controlled by tuning decoherence. The effect of two different unitary noise mechanisms on the evolution of a quantum walk on a linear chain with a generic coin operation is considered in [11]. The first mechanism is a bit-flipchannel noise, restricted to the coin subspace of the quantum walk, and the second is a topological noise caused by randomly broken links in the linear chain. Irrespective of the decoherent dynamics of the walker as a function of the probability per unit time of a decoherent event taking place, the spreading rate is seen to be higher than the corresponding classical rate in both cases. In [12] the authors show that the addition of decoherence to the quantum walk produces a more uniform distribution on the line, and even faster mixing on the cycle by removing the need for time-averaging to obtain a uniform distribution, they calculate numerically the entanglement between the coin and the position of the quantum walker and show that the optimal decoherence rates are such that all the entanglement is just removed by the time the final measurement is made. Quantum walks in multiple dimensions with different quantum coins is studied in [13]. It is shown that by shifting the amplitudes of the coin register in a quantum walk by random phases, we obtain the classical behavior of the quantum walk. In particular, the classical random walk that corresponds to the underlying quantum walk is seen to emerge as a result of the suppression of decoherence. In [14] the authors study continuous-time quantum walks on graphs, focusing on its dynamics under decoherence. In particular they show that decoherence can improve the mixing time in continuous-time quantum walk on cycles. The dynamics of quantum walks on $n$ -dimensional hyper-cycle graph structures is studied in [15]. The authors obtain analytical expressions for the density matrix and probability distribution for weak (quantum) and strong (classical) decoherence modes. The two regimes were shown to have significant differences in how the walking particle spreads throughout the graph. Upper-bound estimate to mixing time was also obtained for both cases. In [16] the authors present a first experimental implementation of a coined discrete time quantum walk and show clear difference with the classical coined quantum walk, since the discrete-time quantum walk possesses destructive interference and periodicity in its evolution. In particular they show that decoherence after each step makes the statistics of the quantum walk tend to that of the classical random walk. Decoherence on quantum random walks over the hypercube is studied in [17]. It is shown that this model possesses a decoherence threshold beneath which the essential properties of the hypercubic quantum walk, such as linear mixing times, are preserved. Beyond the threshold, it is shown that walks behave like their classical counterparts.Two possible routes to classical behavior for the discrete quantum random walk on the line: decoherence in the quantum "coin" which drives the walk, or the use of higher-dimensional coins to dilute the effects of interference is considered in [18]. The authors use the position variance as an indicator of classical behavior, and find analytical expressions for this in the long-time limit. In particular it is observed that the multi-coin walk retains the "quantum" quadratic growth of the variance except in the limit of a new coin for every step, while the walk with decoherence exhibits "classical" linear growth of the variance even for weak decoherence. The quantum walk on a cycle using discrete Wigner functions as a way to represent the states and the evolution of the walker is considered in [19]. It is shown that the use of phase-space representation enables one to develop some intuition about the nature of the decoherence for the quantum walk considered. In particular by coupling the quantum coin to an environment, it is shown that a decoherence model which is roughly equivalent to position diffusion, is obtained.

In [20] neutral atoms trapped in optical lattices to implement quantum random walks on the line and on the circle is considered. The random walk is performed in position space by periodically shifting the lattices and manipulating the internal states of the atom(s)by homogeneous laser pulses. In particular, the influence of decoherence and imperfections in manipulation of internal state of the atoms is investigated. A transition taking place from the ideal quantum random walk to the classical random walk for increasing errors is shown. Quantum walks on a cycle graph, represented by a ring-shape array of quantum dots continuously monitored by individual point contacts, which introduce decoherence is studied in [21]. Analytical expression for the probability distribution along the cycle is obtained for small amount of decoherence. Further it is shown that at fixed low decoherence rates the upper bound estimate for mixing time has linear dependence on the size of the cycle, while fixing the size, one observes inverse linear dependence on the decoherence rate. In [22] quantum walks are considered in a one-dimensional random medium characterized by static or dynamic disorder. It is shown that Anderson localization or decoherence is the main enemy of quantum walks in the presence of static and dynamic disorder, respectively, which destroy their well known quadratic or exponential speed up.

As the literature review shows, many of the previous studies on the effects of decoherence have been experimental in nature due to the complexity of the analytical calculations. Although the analytical calculations have been derived for some particular cases, the general analytic formulas with wide range of applications have been given little attention in the literature. The main aim of this paper of is to give some general analytic expressions for the decoherent quantum walk in two dimensions.

The organization of the paper is as follows. In Section 2 the Kraus operator methodology for studying decoherence on quantum systems is reviewed, the decoherent quantum walk in two dimensions is defined, the Kraus operator representation of the decoherent quantum walk is physically interpreted, the generalized first and second moments for the decoherent quantum walk is obtained, and their Brun formalism is given in Section 3. As an application, in Section 4, we study decoherence in the presence of broken line noise. Section 5 is devoted to the conclusions.

## 2. Analytic Expressions for Moments in the Prescence of General Decoherence

### 2.1. Kraus operator methodology for studying Decoherence on Quantum Systems

According to the authors in [23], the Kraus operator representation is one of widely accepted method of studying the effects of decoherence on quantum systems. For more information on the Kraus operator, the reader is referred to [24]. Define $H_W$ as the Hilbert space of the quantum walker, and let $H_E$ be the Hilbert space of the environment spanned by $\{|e_n\rangle\}_{n=0}^{m}$ where $m$ is the dimension of the environment's Hilbert space.

For the purposes of studying the effects of decoherence on the system, the isolation of the system from the environment is not permissible, therefore according to the authors in [23], we need to find the time evolution of the whole system which consists of the principal system plus the environment, and thereby obtain the state of the whole system by tracing out over the environment's degree of freedom which is given by $\rho_{sys} = Tr_{ENV}\left(U\rho\,{}^{T}\overline{U}\right)$, where $U$ acts on the principal system and the environment Hilbert spaces, $\rho$ is the density operator , $Tr(.)$ denotes the trace operator, the bar denotes conjugation, and in general ${}^{T}A$ denotes the transpose of the matrix $A$. Put $\rho = \rho_0 \otimes |env_0\rangle\langle env_0|$, note that $\rho = \rho_0 \otimes |env_0\rangle\langle env_0|$ gives the state of the whole system, where $|env_0\rangle\langle env_0|$ gives the state of the environment, and $\rho_0$ gives state of the principal system. It follows we can write

$\rho_{sys} = Tr_{ENV}\left(U\rho\,{}^{T}\overline{U}\right)$, as $\rho_{sys} = \sum_{n=0}^{m}\langle e_n|U|env_0\rangle\rho_0\langle env_0|\,{}^{T}\overline{U}|e_n\rangle = \sum_{n=0}^{m}E_n\rho_0\,{}^{T}\overline{E}_n$ , where

$E_n = \langle e_n|U|env_0\rangle$ for $n = 0,1\ldots,m$ are the *Kraus operators*, the Kraus operators satisfy the completeness relation $\sum_{n=0}^{m}E_n\,{}^{T}\overline{E}_n = I$ , that is the operation elements $E_n = \langle e_n|U|env_0\rangle$ are *trace-preserving*

### 2.2 The Decoherent Two Dimensional Quantum Walk

In this section we define the decoherent two dimensional quantum walk . In the two dimensional quantum walk the "coin" degree of freedom is represented by a two-quibit space or coin space, $H_C$, which is spanned by four orthonormal states. To parallel the representation in the one dimensional quantum walk, where $|L\rangle$ and $|R\rangle$ are associated with the left or right displacements respectively, we use the following labeling to represent the four orthonormal states $\{|L\rangle, |R\rangle, |U\rangle, |D\rangle\}$. The position space of the walker, $H_P$ is spanned by the set of orthonormal states $\{|i\rangle, |j\rangle \;\; i, j = -\infty, \ldots \infty\}$. The Hilbert space of the walker is then defined as the tensor product of the coin space $H_C$ and the position space $H_P$, that is, if $H'$ is the whole Hilbert space of the walker, then $H' = H_p \otimes H_c$. To define the movement of the quantum walker in two dimensions, we first consider what happens on one step in the quantum walk. We first make superposition on the coin space with the coin operator $U_C$ and move the particle according to the coin state with the translation operator $S$ as follows $U_w = S \cdot (I \otimes U_c)$, where $I$ is the identity operator in $H_P$, $U_W$ is the coin operator on the position space, and the translation operator $S$ is given by

$$S = \sum_{x,y} \{ |x-1, y\rangle\langle x, y| \otimes |L\rangle\langle L| + |x+1, y\rangle\langle x, y| \otimes |R\rangle\langle R| + |x, y-1\rangle\langle x, y| \otimes |D\rangle\langle D| + |x, y+1\rangle\langle x, y| \otimes |U\rangle\langle U| \}$$

The evolution of the two dimensional quantum walk is then defined by $|\Psi(t+1)\rangle = U_w |\Psi(t)\rangle$. By induction on $t$, we can show that evolution of the two dimensional quantum walk in terms of the initial state is given by $|\Psi(t+1)\rangle = U_w{}^t |\Psi(0)\rangle$. Since the time evolution of the quantum walker in the presence of decoherence is not pure, the time evolution consist of the Hilbert space of the walker and that of the environment which is given by the following tensor product $H'' = H_E \otimes H'$, where $H_E$ is the Hilbert space of the environment and $H'$ is as previously defined. The Kraus operator representation of the evolution of the walk is defined by $\rho(t+1) = \sum_{n=0}^{m} E_n \, \rho(t) \, {}^T\overline{E}_n$. By induction on $t$, we can show that in terms of the initial state of the whole system the Kraus operator representation of the evolution of the walk is defined by $\rho(t) = \left( \sum_{n=0}^{m} \right)^t E_n{}^t \rho(0) \left( {}^T\overline{E}_n \right)^t$. In order to obtain the final state of $\rho(t)$ we need to find the Kraus operators for

$$\rho_{sys} = \sum_{n=0}^{m} \langle e_n | U | env_0 \rangle \rho_0 \langle env_0 | {}^T\overline{U} | e_n \rangle = \sum_{n=0}^{m} E_n \rho_0 \, {}^T\overline{E}_n$$ , where $E_n = \langle e_n | U | env_0 \rangle$ for

$n = 0,1\ldots,m$. Since the Kraus operators act on the walker's Hilbert space, it follows that we can write

$$E_n = \sum_{x,y,x',y'} \sum_{i,j,k,l} a^{(n)}_{x,y,x',y',i,j,k,l} \, |x', y'\rangle \langle x, y| \otimes |i\rangle\langle j| |k\rangle\langle l|$$

$$= \sum_{x,y} \sum_{m,n} \sum_{i,j,k,l} a^{(n)}_{x,y,m,n,i,j,k,l} \, |x+m, y+n\rangle \langle x, y| \otimes |i\rangle\langle j| |k\rangle\langle l|$$

where $x, y, m, n = -\infty,\ldots,\infty$ and $i, j, k, l = \{L, R, D, U\}$.

### 2.3 Kraus Operator Representation of Impure Evolution: Physical Interpretation

Suppose $p_i$ is the probability that the $ith$ unknown reason affects the state of the system. Let $A_i$ be the operator that evolves the state of the system whenever the $ith$ reason affects it. Since the environment is dynamic, it follows that there are infinitely many ways the state of the system could be affected. Without loss of generality, we may assume that whenever the $ith$ state of the environment is $|e_i\rangle$, the operator $A_i$ evolves the state with probability $p_i$ for $i = 1, 2, \ldots, r$, so then we can assume that $H_E = span\{|e_1\rangle,\ldots,|e_r\rangle\}$, where the basis consist of orthonormal vectors. Since the initial state of the environment $|env_0\rangle \in H_E$, it follows we can write $|env_0\rangle$ as a linear combination of the basis vectors, say, $|env_0\rangle = c_1|e_1\rangle + c_2|e_2\rangle + \ldots + c_{r-1}|e_{r-1}\rangle + c_r|e_r\rangle$, where $c_1,\ldots c_{r-1}, c_r$ are constants to be determined. However from $|env_0\rangle = c_1|e_1\rangle + c_2|e_2\rangle + \ldots + c_{r-1}|e_{r-1}\rangle + c_r|e_r\rangle$, one sees that $c_i^2$ is the probability of finding the environment in state $|e_i\rangle$, however, the probability of finding the environment in state $|e_i\rangle$ is $p_i$, it then follows that $c_i^2 = p_i$, and so $c_i = \sqrt{p_i}$ are the constants, and so $|env_0\rangle = \sqrt{p_1}|e_1\rangle + \sqrt{p_2}|e_2\rangle + \ldots + \sqrt{p_{r-1}}|e_{r-1}\rangle + \sqrt{p_r}|e_r\rangle$. Since in the presence of decoherence

the Hilbert space is given by $H'' = H_E \otimes H'$, it follows that the unitary transformation of the whole system can be written as $U = \sum_{i=1}^{r} |e_i\rangle\langle e_i| \otimes A_i$ . Since the basis vectors are orthonormal, it follows we can write $\langle e_i|U|env_0\rangle = \langle e_i|\{|e_i\rangle\langle e_i| \otimes A_i\}\sqrt{p_i}|e_i\rangle = \sqrt{p_i}\, A_i$, but the definition of the Kraus operators imply, the Kraus operators are given by $E_i = \sqrt{p_i}\, A_i$. Recall the following Kraus operator representation $\rho(t+1) = \sum_{n=0}^{m} E_n\, \rho(t)\, {}^T\overline{E}_n$ for the walk. It follows from it that the evolution of the walk can now be written in terms of a new density operator $\rho'$ defined as $\rho' = \sum_{i=1}^{r} E_i\, \rho\, {}^T\overline{E}_i$, where $E_i = \sqrt{p_i}\, A_i$. Substituting, $E_i = \sqrt{p_i}\, A_i$ in $\rho' = \sum_{i=1}^{r} E_i\, \rho\, {}^T\overline{E}_i$, it follows we can write $\rho' = \sum_{i=1}^{r} p_i A_i\, \rho\, {}^T\overline{A}_i$ . Using $\rho' = \sum_{i=1}^{r} p_i A_i\, \rho\, {}^T\overline{A}_i$ in conjunction with $|env_0\rangle = \sqrt{p_1}|e_1\rangle + \sqrt{p_2}|e_2\rangle + \ldots + \sqrt{p_{r-1}}|e_{r-1}\rangle + \sqrt{p_r}|e_r\rangle$ and $U = \sum_{i=1}^{r} |e_i\rangle\langle e_i| \otimes A_i$, the Kraus operator representation for the impure evolution can be given.

### 2.4 Brun Formalism of the Moments for the Decoherent Quantum Walk

In this section we obtain the generalized first and second moments of the two dimensional quantum walk in the presence of decoherence. We will use the Fourier transformation approach as used by the authors [23] in the case of obtaining the Brun formalism of the generalized moments for the decoherent quantum walk in one dimension. In the two dimensional case the Fourier transformation is defined as follows $|x, y\rangle = \int_{-\pi}^{\pi}\int_{-\pi}^{\pi} \frac{dk_x dk_y}{4\pi^2}\, e^{-ik_x x - k_y y}\, |k_x, k_y\rangle$. Using this transformation, it follows that we can write

$$E_n = \sum_{x,y,x',y'} \sum_{m,n} \sum_{i,j,k,l} a^{(n)}_{x,y,x',y',i,j,k,l} \left| x+m, y+n \right\rangle \left\langle x,y \right| \otimes \left| i \right\rangle \left\langle j \right| \left| k \right\rangle \left\langle l \right|$$ in the $(k_x, k_y)$-space as

$$\tilde{E}_n = \sum_{x,y,x',y'} \sum_{m,n} \sum_{i,j,k,l} a^{(n)}_{x,y,x',y',i,j,k,l} \iiiint \frac{dk_x dk_y dk'_x dk'_y}{16\pi^4} e^{-i(k_x m + k_y n)} e^{-ix(k_x - k'_x) - iy(k_y - k'_y)} \left| k_x, k_y \right\rangle \left\langle k'_x, k'_y \right| \otimes \left| i \right\rangle \left\langle j \right| \left| k \right\rangle \left\langle l \right|$$

Assuming that the coefficients $a^{(n)}_{x,y,x',y',i,j,k,l}$ are not dependent on the coordinate $(x, y)$, then the probability of translation in the position space depends on lengths $m, n$ of the translation in the $x-axis$ and $y-axis$ respectively, but not on the position $(x, y)$ where the translation occurs. Under this assumption we can write $\tilde{E}_n$ as

$$\tilde{E}_n = \sum_{m,n} \sum_{i,j,k,l} a^{(n)}_{m,n,i,j,k,l} \iiiint \frac{dk_x dk_y dk'_x dk'_y}{4\pi^2} e^{-i(k_x m + k_y n)} \delta(k_x - k'_x) \delta(k_y - k'_y) \left| k_x, k_y \right\rangle \left\langle k'_x, k'_y \right| \otimes \left| i \right\rangle \left\langle j \right| \left| k \right\rangle \left\langle l \right|$$

where the orthonormalization relations $\sum_x e^{-ix(k_x - k'_x)} = 2\pi \delta(k_x - k'_x)$ and $\sum_y e^{-iy(k_y - k'_y)} = 2\pi \delta(k_y - k'_y)$ have been used. Note that in these relations $\delta(\cdot)$ is the Dirac Delta. By integrating with respect to $k'_x$ and $k'_x$ respectively and changing the order of integration and summation we get

$$\tilde{E}_n = \iint \frac{dk_x dk_y}{4\pi^2} \left| k_x, k_y \right\rangle \left\langle k_x, k_y \right| \otimes C_n(k_x, k_y)$$ , where

$C_n(k_x, k_y) = \sum_{m,n} \sum_{i,j,k,l} a^{(n)}_{m,n,i,j,k,l} e^{-i(k_x m + k_y n)} \left| i \right\rangle \left\langle j \right| \left| k \right\rangle \left\langle l \right|$. Now writing the general form of $\rho_0$ in the $(k_x, k_y)$-space as $\rho_0 = \iint \iint \frac{dk_x dk_y dk'_x dk'_y}{16\pi^4} \left| k_x, k_y \right\rangle \left\langle k'_x, k'_y \right| \otimes \left| \psi_0 \right\rangle \left\langle \psi_0 \right|$, then from the Kraus operator representation of the walk, the first step is given by

$$\rho' = \iint \iint \frac{dk_x dk_y dk'_x dk'_y}{16\pi^2} \left| k_x, k_y \right\rangle \left\langle k'_x, k'_y \right| \otimes \sum_n C_n(k_x, k_y) |\psi_0\rangle\langle\psi_0| \,^T\overline{C}_n(k'_x, k'_y)$$

$$= \iint \iint \frac{dk_x dk_y dk'_x dk'_y}{16\pi^2} \left| k_x, k_y \right\rangle \left\langle k'_x, k'_y \right| \otimes L_{k_x,k_y,k'_x,k'_y} |\psi_0\rangle\langle\psi_0|$$

where the superoperator, $L_{k_x,k_y,k'_x,k'_y}$, is defined by $L_{k_x,k_y,k'_x,k'_y} \widetilde{O} = \sum_n C_n(k_x, k_y) \widetilde{O} \,^T\overline{C}_n(k'_x, k'_y)$

In terms of the superoperator, by induction on $t$, we can show that the Kraus operator representation gives the evolution of the walk after $t$ steps as

$\rho(t) = \iint \iint \frac{dk_x dk_y dk'_x dk'_y}{16\pi^4} \left| k_x, k_y \right\rangle \left\langle k'_x, k'_y \right| \otimes L^t{}_{k_x,k_y,k'_x,k'_y} |\psi_0\rangle\langle\psi_0|$. The probability of findng the walker in position $(x, y)$ is given by

$$p(x, y, t) = \int\int\int\int \frac{dk_x dk_y dk'_x dk'_y}{16\pi^4} \langle x, y | k_x, k_y \rangle \langle k'_x, k'_y | x, y \rangle Tr\left(L^t{}_{k_x,k_y,k'_x,k'_y}\, \rho_0\right)$$

$$= \int\int\int\int \frac{dk_x dk_y dk'_x dk'_y}{16\pi^4} e^{-ix(k'_x - k_x) - iy(k'_y - k_y)}\, Tr\left(L^t{}_{k_x,k_y,k'_x,k'_y}\, \rho_0\right) \tag{1}$$

Since the Kraus operators satisfy the so-called completeness relations, we also have $\sum_n {}^T\overline{C}_n(k_x, k_y) C_n(k_x, k_y) = I$. The condition on the coin operator imply that the super-operators $L_{k_x,k_y,k_x,k_y}$ are *trace preserving*. In particular by induction on $m$ we can show that $Tr\left(L^m{}_{k_x,k_y,k_x,k_y} \widetilde{O}\right) = Tr\left(\widetilde{O}\right)$ for any arbitrary operator $\widetilde{O}$. In order to know the state and probability distribution of the system, we need to know the operators $C_n(k_x, k_y)$ and thus use $p(x, y, t)$ and $p(t)$ immediately above, however, this information is lacking thus we look to the moments of the distribution. The $mth$ moment of the probability distribution $p(x, y, t)$ is defined by

$$\left\langle (xy)^m \right\rangle = \sum_{x,y} (xy)^m\, p(x, y, t) \tag{2}$$

Now inserting (1) into (2) we get

$$\left\langle (xy)^m \right\rangle = \sum_{x,y} (xy)^m \iiiint \frac{dk_x dk_y dk'_x dk'_y}{16\pi^4} e^{-ix(k'_x - k_x) - iy(k'_y - k_y)} Tr\left(L^t{}_{k_x,k_y,k'_x,k'_y}\ \rho_0\right) \tag{3}$$

Now take $m=1$ in equation (3) above we get

$$\left\langle (xy) \right\rangle = \sum_{x,y} (xy) \iiiint \frac{dk_x dk_y dk'_x dk'_y}{16\pi^4} e^{-ix(k'_x - k_x) - iy(k'_y - k_y)} Tr\left(L^t{}_{k_x,k_y,k'_x,k'_y}\ \rho_0\right) \tag{4}$$

However, the orthonormalization relations imply we can write $\sum_x e^{ix(k'_x - k_x)} = 2\pi\,\delta(k'_x - k_x)$ and

$\sum_y e^{iy(k'_y - k_y)} = 2\pi\,\delta(k'_y - k_y)$. Now differentiating with respect to $k_x$ and $k_y$ respectively we get

$$\sum_x x e^{ix(k'_x - k_x)} = -2\pi i \frac{d}{dk_x}\delta(k'_x - k_x) \text{ and } \sum_y y e^{iy(k'_y - k_y)} = -2\pi i \frac{d}{dk_y}\delta(k'_y - k_y) \tag{5}$$

Now substitute (5) into (4) we get

$$\left\langle (xy) \right\rangle = \iiiint \frac{dk_x dk_y dk'_x dk'_y}{-4\pi^2} \frac{d}{dk_x}\delta(k'_x - k_x) \frac{d}{dk_y}\delta(k'_y - k_y)\ Tr\left(L^t{}_{k_x,k_y,k'_x,k'_y}\ \rho_0\right) \tag{6}$$

Now take $m=2$ in equation (3) we get

$$\left\langle (xy)^2 \right\rangle = \sum_{x,y} (xy)^2 \iiiint \frac{dk_x dk_y dk'_x dk'_y}{16\pi^4} e^{-ix(k'_x - k_x) - iy(k'_y - k_y)} Tr\left(L^t{}_{k_x,k_y,k'_x,k'_y}\ \rho_0\right) \tag{7}$$

Now differentiate the first and second expression in (5) with respect to $k'_x$ and $k'_y$ respectively we get

$$\sum_x x^2 e^{ix(k'_x - k_x)} = -2\pi \frac{d^2}{dk'_x dk_x}\delta(k'_x - k_x) \text{ and } \sum_y y^2 e^{iy(k'_y - k_y)} = -2\pi \frac{d^2}{dk'_y dk_y}\delta(k'_y - k_y) \tag{8}$$

Now put the expressions in equation (8) into equation (7) we get

$$\left\langle (xy)^2 \right\rangle = \iiiint \frac{dk_x dk_y dk'_x dk'_y}{4\pi^2} \frac{d^2}{dk'_x dk_x}\delta(k'_x - k_x) \frac{d^2}{dk'_y dk_y}\delta(k'_y - k_y) Tr\left(L^t{}_{k_x,k_y,k'_x,k'_y}\ \rho_0\right) \tag{9}$$

In order to carry out the integrations in (6) and (9) we need the following equations

$$\frac{d}{dk_x} Tr\left(L_{k_x,k_y,k'_x,k'_y}\,\tilde{O}\right) = Tr\left(\sum_n \frac{d}{dk_x} C_n\left(k_x,k_y\right)\tilde{O}\ {}^T\overline{C}_n(k'_x,k'_y)\right) \tag{10}$$

$$\frac{d}{dk_y} Tr\left(L_{k_x,k_y,k'_x,k'_y}\,\tilde{O}\right) = Tr\left(\sum_n \frac{d}{dk_y} C_n\left(k_x,k_y\right)\tilde{O}\ {}^T\overline{C}_n(k'_x,k'_y)\right) \tag{11}$$

$$\frac{d}{dk'_x} Tr\left(L_{k_x,k_y,k'_x,k'_y}\,\tilde{O}\right) = Tr\left(\sum_n C_n\left(k_x,k_y\right)\tilde{O}\ \frac{d}{dk'_x}\ {}^T\overline{C}_n(k'_x,k'_y)\right) \tag{12}$$

$$\frac{d}{dk'_y} Tr\left(L_{k_x,k_y,k'_x,k'_y}\,\tilde{O}\right) = Tr\left(\sum_n C_n\left(k_x,k_y\right)\tilde{O}\ \frac{d}{dk'_y}\ {}^T\overline{C}_n(k'_x,k'_y)\right) \tag{13}$$

Notice from $C_n(k_x,k_y) = \sum_{m,n}\sum_{i,j,k,l} a^{(n)}_{m,n,i,j,k,l} e^{-i(k_x m + k_y n)} |i\rangle\langle j||k\rangle\langle l|$, that we also have the following

derivatives related to (10)-(13).

$$\frac{dC_n(k_x,k_y)}{dk_x} = -i\sum_{m,n}\sum_{i,j,k,l} m\, a^{(n)}_{m,n,i,j,k,l} e^{-i(k_x m + k_y n)} |i\rangle\langle j||k\rangle\langle l| \tag{14}$$

$$\frac{dC_n(k_x,k_y)}{dk_y} = -i\sum_{m,n}\sum_{i,j,k,l} n\, a^{(n)}_{m,n,i,j,k,l} e^{-i(k_x m + k_y n)} |i\rangle\langle j||k\rangle\langle l| \tag{15}$$

Now agreeing to write ${}^T\overline{A} = A^*$, it follows that we have the following

$$\frac{d\ {}^T\overline{C}_n(k'_x,k'_y)}{dk'_x} = i\sum_{m,n}\sum_{i,j,k,l} m\left(a^{(n)}_{m,n,i,j,k,l}\right)^* e^{i(k'_x m + k'_y n)} |l\rangle\langle k||j\rangle\langle i| \tag{16}$$

$$\frac{d\ {}^T\overline{C}_n(k'_x,k'_y)}{dk'_y} = i\sum_{m,n}\sum_{i,j,k,l} n\left(a^{(n)}_{m,n,i,j,k,l}\right)^* e^{i(k'_x m + k'_y n)} |l\rangle\langle k||j\rangle\langle i| \tag{17}$$

Since the superoperator $L_{k_x,k_y,k'_x,k'_y}$ acts on the density matrix which is positive and Hermitian, we can

in this case write equations (10)-(13) as follows

$$\frac{d}{dk_x} Tr\left(L_{k_x,k_y,k'_x,k'_y}\,\rho\right) = Tr\left(G_{k_x,k_y,k'_x,k'_y}\,\rho\right) \tag{18}$$

$$\frac{d}{dk_y} Tr\left(L_{k_x,k_y,k'_x,k'_y}\,\rho\right) = Tr\left(G_{k_x,k_y,k'_x,k'_y}\,\rho\right) \tag{19}$$

$$\frac{d}{dk'_x} Tr\left(L_{k_x,k_y,k'_x,k'_y}\,\rho\right) = Tr\left({}^{T}\overline{G}_{k'_y,k'_x,k_y,k_x}\,\rho\right) \tag{20}$$

$$\frac{d}{dk'_y} Tr\left(L_{k_x,k_y,k'_x,k'_y}\,\rho\right) = Tr\left({}^{T}\overline{G}_{k'_y,k'_x,k_y,k_x}\,\rho\right) \tag{21}$$

where we have used in (18)-(21) the following

$$G_{k_x,k_y,k'_x,k'_y}\tilde{O} = \sum_n \frac{d^2}{dk_x dk_y} C_n\left(k_x,k_y\right)\tilde{O}\ {}^{T}\overline{C}_n\left(k'_x,k'_y\right) \tag{22}$$

To carry out the integration in equation (6), we use the method of integration by parts, first we notice that we can write equation (6) as

$$\left\langle\,(xy)\right\rangle = \frac{-1}{2\pi}\int\!\!\int dk_x dk_y \left\{ \frac{1}{2\pi}\iint dk'_x dk'_y \frac{d}{dk_x}\delta(k'_x - k_x)\frac{d}{dk_y}\delta(k'_y - k_y)\; Tr\left(L^{t}{}_{k_x,k_y,k'_x,k'_y}\;\rho_0\right)\right\} \tag{23}$$

We notice that the superoperator inside the trace function is a power of $t$, it follows that equation (23) is a function of $t$, let us agree to write this function as

$$\left\langle\,(xy)\right\rangle_t = \frac{-1}{2\pi}\int\!\!\int dk_x dk_y \left\{ \frac{1}{2\pi}\iint dk'_x dk'_y \frac{d}{dk_x}\delta(k'_x - k_x)\frac{d}{dk_y}\delta(k'_y - k_y)\; Tr\left(L^{t}{}_{k_x,k_y,k'_x,k'_y}\;\rho_0\right)\right\} \tag{24}$$

If $t = 1$, agree to write $\left\langle\,(xy)\right\rangle_1 = \left\langle (xy)\right\rangle$, then equation (24) implies that

$$\left\langle\,(xy)\right\rangle = \frac{-1}{2\pi}\int\!\!\int dk_x dk_y \left\{ \frac{1}{2\pi}\iint dk'_x dk'_y \frac{d}{dk_x}\delta(k'_x - k_x)\frac{d}{dk_y}\delta(k'_y - k_y)\; Tr\left(L_{k_x,k_y,k'_x,k'_y}\;\rho_0\right)\right\} \tag{25}$$

The double integral inside the parenthesis of the expression in equation (25) can be written as follows

$$\frac{1}{2\pi}\iint dk'_x dk'_y \frac{d^2}{dk_x dk_y}\delta(k'_x - k_x, k'_y - k_y)\, Tr\left(L_{k_x,k_y,k'_x,k'_y}\, \rho_0\right) \tag{26}$$

where $\delta(\cdot,\cdot)$ denotes the two-dimensional Dirac Delta function.

Now using (26) inside (25) and carrying out the integration, the whole of (25) becomes

$$\frac{-1}{2\pi}\left\{\int_{-\pi}^{\pi}\int_{-\pi}^{\pi}\frac{dk_x dk_y}{2\pi} Tr\left({}^T\overline{G}_{k_x,k_y}\left(G_{k_x,k_y}\rho_0\right) + G_{k_x,k_y}\left({}^T\overline{G}_{k_x,k_y}\rho_0\right)\right) + \int_{-\pi}^{\pi}\int_{-\pi}^{\pi}\frac{dk_x dk_y}{2\pi} Tr\left(\Omega_{k_x,k_y}\rho_0\right)\right\} \tag{27}$$

where we have defined

$$\Omega_{k_x,k_y} = \left.\frac{d^2\ {}^T\overline{G}_{k_x,k_y,k'_x,k'_y}}{dk_x dk_y}\right|_{k_x=k'_x,\,k_y=k'_y} = \sum_n \frac{d^2 C_n(k_x,k_y)}{dk_x dk_y}\tilde{O}\left.\frac{d^2\ {}^T\overline{C}_n(k'_x,k'_y)}{dk'_x dk'_y}\right|_{k_x=k'_x,\,k_y=k'_y} \tag{28}$$

and we have also used $G_{k_x,k_y} \equiv G_{k_x,k_y,k_x,k_y}$ and $L_{k_x,k_y} \equiv L_{k_x,k_y,k_x,k_y}$. In general, by induction on $t$, the first moment can be shown to take the following form

$$\langle (xy)\rangle_t = \frac{-1}{2\pi}\int_{-\pi}^{\pi}\int_{-\pi}^{\pi}\frac{dk_x dk_y}{2\pi}\sum_{j=1}^{t}\sum_{k=1}^{j-1} Tr\left\{{}^T\overline{G}_{k_x,k_y} L^{j-k-1}{}_{k_x,k_y}\left(G_{k_x,k_y} L^{k-1}{}_{k_x,k_y}\rho_0\right) + G_{k_x,k_y} L^{j-k-1}{}_{k_x,k_y}\left({}^T\overline{G}_{k_x,k_y} L^{k-1}{}_{k_x,k_y}\rho_0\right)\right\}$$
$$+\frac{-1}{2\pi}\int_{-\pi}^{\pi}\int_{-\pi}^{\pi}\frac{dk_x dk_y}{2\pi}\sum_{j=1}^{t} Tr\left(\Omega_{k_x,k_y}\left(L^{j-1}{}_{k_x,k_y}\rho_0\right)\right) \tag{29}$$

where $\Omega_{k_x,k_y}$, $G_{k_x,k_y}$, $L_{k_x,k_y}$ take the same definitions as in equation (28).

Now to obtain the second moment in a form similar to equation (29) we notice we can write equation (9) as

$$\left\langle (xy)^2\right\rangle_t = \frac{1}{2\pi}\iint\frac{dk_x dk_y}{2\pi}\left\{\frac{1}{2\pi}\iint dk'_x dk'_y \frac{d^4\,\delta(k'_x - k_x, k'_y - k_y)}{dk_x dk'_x dk'_y dk_y} Tr\left(L^t{}_{k_x,k_y,k'_x,k'_y}\rho_0\right)\right\} \tag{30}$$

where we have written the second moment as a function of $t$, since the superoperator is a power of $t$ in equation (9). Further inspection of equation (30) implies we can write the second moment as

$$\left\langle (xy)^2 \right\rangle_t = \frac{1}{2\pi}\iint \frac{dk_x dk_y}{2\pi}\left\{\frac{d^2}{dk'_x dk'_y}\left\{\frac{1}{2\pi}\iint dk'_x dk'_y \frac{d^2\,\delta(k'_x - k_x, k'_y - k_y)}{dk_x dk_y} Tr\left(L^t{}_{k_x,k_y,k'_x,k'_y}\,\rho_0\right)\right\}\right\} \qquad (31)$$

Equation (31) can further be written as

$$\left\langle (xy)^2 \right\rangle_t = \frac{d^2}{dk'_x dk'_y}\left\{\frac{1}{2\pi}\iint \frac{dk_x dk_y}{2\pi}\left\{\frac{1}{2\pi}\iint dk'_x dk'_y \frac{d^2\,\delta(k'_x - k_x, k'_y - k_y)}{dk_x dk_y} Tr\left(L^t{}_{k_x,k_y,k'_x,k'_y}\,\rho_0\right)\right\}\right\} \qquad (32)$$

From which it follows that

$$\left\langle (xy)^2 \right\rangle_t = -\frac{d^2}{dk'_x dk'_y}\left\langle (xy) \right\rangle_t \ , \qquad (33)$$

where $\left\langle (xy) \right\rangle_t$ is given by equation (29).

We will not carry out the differentiation explicitly, however we note that equation (33) is of the Brun-type if and only if the first moment is of the Brun-type, since the explicit differentiation only comes from the trace terms in the first moment.

### 3. Coin Decoherence

In this section we attempt to show that in the coin decoherence, the general formalism for the first and second moments obtained in the previous section are of the Brun-type. As we mentioned it is necessary only to show that the first moment is of the Brun-type.

If we assume that the operators $D_c^{(n)}$ act on the coin subspace with probability $p_n$, then recalling the physical interpretation of the Kraus operator representation of the impure evolution in Section 2.3, whenever the environment is in state $\left| e_n \right\rangle$ the operator $A_n$ acts on the systems Hilbert space with probability $p_n$, in particular the unitary transformation of the whole system is given by

$$\sum_{n=1}^{r} \left| e_n \right\rangle \left\langle e_n \right| \otimes A_n .$$

In light of the operator $D_c^{(n)}$ acting on the coin subspace, in one step of the quantum walk, whenever the system is in state $|e_n\rangle$, we first make superposition on the coin space of the systems Hilbert space with the operator $H''D_c^{(n)}$ and after that we move the particle according to the coin state with the translation operator S as follows $A_n = S\cdot\left(I\otimes H''D_c^{(n)}\right)$, where

$$S=\sum_{x,y}\left\{|x-1,y\rangle\langle x,y|\otimes|L\rangle\langle L|+|x+1,y\rangle\langle x,y|\otimes|R\rangle\langle R|+|x,y-1\rangle\langle x,y|\otimes|D\rangle\langle D|+|x,y+1\rangle\langle x,y|\otimes|U\rangle\langle U|\right\}.$$

However, we recall from Section 2.3 that in the presence of decoherence the Kraus operators take the form $E_n=\sqrt{p_n}A_n$. Using $A_n = S\cdot\left(I\otimes H''D_c^{(n)}\right)$ in $E_n=\sqrt{p_n}A_n$ with the explicit form of $S$, we get

$$E_n=\sqrt{p_n}\sum_{x,y}\left\{\begin{array}{l}|x-1,y\rangle\langle x,y|\otimes|L\rangle\langle L|HD_c^{(n)}+|x+1,y\rangle\langle x,y|\otimes|R\rangle\langle R|HD_c^{(n)}+|x,y-1\rangle\langle x,y|\otimes|D\rangle\langle D|HD_c^{(n)}\\+|x,y+1\rangle\langle x,y|\otimes|U\rangle\langle U|HD_c^{(n)}\end{array}\right\}\quad(34)$$

Since the basis for the coin space is $\{|L\rangle,|R\rangle,|D\rangle,|U\rangle\}$ we can write the operator $H''D_c^{(n)}$ as a linear combination of the basis vectors, say $H''D_c^{(n)}=\sum_{p,q,r,s\in\{L,R,D,U\}}\gamma_{p,q,r,s}^{(n)}|p\rangle\langle q\|r\rangle\langle s|$, then, equation (34) becomes

$$E_n=\sqrt{p_n}\sum_{x,y}\sum_{q,r,s\in\{R,D,U\}}\left\{\begin{array}{l}\gamma^{(n)}{}_{L,q,r,s}|x-1,y\rangle\langle x,y|\otimes|L\rangle\langle q\|r\rangle\langle s|+\gamma^{(n)}{}_{R,q,r,s}|x+1,y\rangle\langle x,y|\otimes|R\rangle\langle q\|r\rangle\langle s|+\\\gamma^{(n)}{}_{D,q,r,s}|x,y-1\rangle\langle x,y|\otimes|D\rangle\langle q\|r\rangle\langle s|+\gamma^{(n)}{}_{U,q,r,s}|x,y+1\rangle\langle x,y|\otimes|U\rangle\langle q\|r\rangle\langle s|\end{array}\right\}\quad(35)$$

Upon inspection one sees that equation (35) is similar in nature to the following

$E_n=\sum_{x,y}\sum_{m,n}\sum_{i,j,k,l}a_{x,y,m.n,i,j,k,l}^{(n)}\,|x+m,y+n\rangle\,\langle x,y|\otimes|i\rangle\langle j||k\rangle\langle l|$, which we encountered earlier on, with the specified values of $m=-1,1,0$ and $n=-1,1,0$, thus we have

$a^{(n)}_{-1,0,L,q,r,s} = \gamma^{(n)}_{L,q,r,s}$, $a^{(n)}_{1,0,R,q,r,s} = \gamma^{(n)}_{R,q,r,s}$, $a^{(n)}_{0,-1,D,q,r,s} = \gamma^{(n)}_{D,q,r,s}$, $a^{(n)}_{0,1,U,q,r,s} = \gamma^{(n)}_{U,q,r,s}$. We can therefore

write $C_n(k_x, k_y) = \sum_{m,n} \sum_{i,j,k,l} a^{(n)}_{m,n,i,j,k,l} e^{-i(k_x m + k_y n)} |i\rangle\langle j||k\rangle\langle l|$ as

$$C_n(k_x, k_y) = \sum_{j,k,l=\{L,R,D,U\}} \left\{ \begin{matrix} \gamma^{(n)}_{L,q,r,s}\, e^{ik_x} |L\rangle\langle j\|k\rangle\langle l| + \gamma^{(n)}_{R,q,r,s}\, e^{-ik_x} |R\rangle\langle j\|k\rangle\langle l| + \gamma^{(n)}_{D,q,r,s}\, e^{ik_y} |D\rangle\langle j\|k\rangle\langle l| + \\ \gamma^{(n)}_{U,q,r,s}\, e^{-ik_y} |U\rangle\langle j\|k\rangle\langle l| \end{matrix} \right\} \tag{36}$$

From equation (36) we notice that $\frac{d^2 C_n(k_x, k_y)}{dk_x dk_y} = Z C_n(k_x, k_y)$. The definition of $L_{k_x,k_y,k'_x,k'_x}$ and

$G_{k_x,k_y,k'_x,k'_y} \widetilde{O} = \sum_n \frac{d^2}{dk_x dk_y} C_n(k_x, k_y) \widetilde{O}\ {}^T\overline{C}_n(k'_x, k'_y)$, imply that

$G_{k_x,k_y,k'_x,k'_y} \widetilde{O} = Z\left(L_{k_x,k_y,k'_x,k'_y} \widetilde{O}\right)$ and ${}^T\overline{G}_{k_x,k_y,k'_x,k'_y} \widetilde{O} = \left(L_{k_x,k_y,k'_x,k'_y} \widetilde{O}\right) Z$. On the other hand $\Omega_{k_x,k_y}$ in

equation (28) becomes $\Omega_{k_x k_y} \widetilde{O} = \left. \frac{d^2 G_{k_x,k_y,k'_x,k'_y} \widetilde{O}}{dk_x dk_y} \right|_{k_x = k'_x, k_y = k'_y} = Z\left(L_{k_x,k_y} \widetilde{O}\right) Z$. Now to see that the

first moment expression obtained is of the Brun-type, notice that upon using

$\Omega_{k_x k_y} \widetilde{O} = \left. \frac{d^2 G_{k_x,k_y,k'_x,k'_y} \widetilde{O}}{dk_x dk_y} \right|_{k_x = k'_x, k_y = k'_y} = Z\left(L_{k_x,k_y} \widetilde{O}\right) Z$, the last term on the right in equation (29) can

be written as

$$\frac{-1}{2\pi} \int_{-\pi}^{\pi}\int_{-\pi}^{\pi} \frac{dk_x dk_y}{2\pi} \sum_{j=1}^{t} Tr\left(\Omega_{k_x,k_y}\left(L^{j-1}{}_{k_x,k_y} \rho_0\right)\right) = \frac{-1}{2\pi} \int_{-\pi}^{\pi}\int_{-\pi}^{\pi} \frac{dk_x dk_y}{2\pi} \sum_{j=1}^{t} Tr\left(L^{j}{}_{k_x,k_y} \rho_0\right) \tag{37}$$

Now the linearity of the trace function implies we can write the first term on the right in equation

(29) as

$$\frac{-1}{2\pi} \int_{-\pi}^{\pi}\int_{-\pi}^{\pi} \frac{dk_x dk_y}{2\pi} \sum_{j=1}^{t} \sum_{k=1}^{j-1} Tr\left\{ {}^T\overline{G}_{k_x,k_y} L^{j-k-1}{}_{k_x,k_y} \left(G_{k_x,k_y} L^{k-1}{}_{k_x,k_y} \rho_0\right) \right\} + Tr\left\{ G_{k_x,k_y} L^{j-k-1}{}_{k_x,k_y} \left({}^T\overline{G}_{k_x,k_y} L^{k-1}{}_{k_x,k_y} \rho_0\right) \right\} \tag{38}$$

Now using $G_{k_x,k_y,k'_x,k'_y}\tilde{O} = Z\left(L_{k_x,k_y,k'_x,k'_y}\tilde{O}\right)$ and ${}^{T}\overline{G}_{k_x,k_y,k'_x,k'_y}\tilde{O} = \left(L_{k_x,k_y,k'_x,k'_y}\tilde{O}\right)Z$ , the first trace term in equation (38) can be written as

$$Tr\left\{{}^{T}\overline{G}_{k_x,k_y} L^{j-k-1}{}_{k_x,k_y}\left(G_{k_x,k_y} L^{k-1}{}_{k_x,k_y}\rho_0\right)\right\} = Tr\left\{Z\,L^{j-k}{}_{k_x,k_y}\left(ZL^{k}{}_{k_x,k_y}\rho_0\right)\right\} \tag{39}$$

Similarly, the second trace term can be written as

$$Tr\left\{G_{k_x,k_y} L^{j-k-1}{}_{k_x,k_y}\left({}^{T}\overline{G}_{k_x,k_y} L^{k-1}{}_{k_x,k_y}\rho_0\right)\right\} = Tr\left\{Z\,L^{j-k}{}_{k_x,k_y}\left(L^{k}{}_{k_x,k_y}\rho_0\right)Z\right\} \tag{40}$$

Putting (39) and (40) into (38) the first term of (29) can be written as

$$\frac{-1}{2\pi}\int_{-\pi}^{\pi}\int_{-\pi}^{\pi}\frac{dk_x dk_y}{2\pi}\sum_{j=1}^{t}\sum_{k=1}^{j-1} Tr\left\{Z\,L^{j-k}{}_{k_x,k_y}\left(ZL^{k}{}_{k_x,k_y}\rho_0\right)\right\} + Tr\left\{Z\,L^{j-k}{}_{k_x,k_y}\left(L^{k}{}_{k_x,k_y}\rho_0\right)Z\right\} \tag{41}$$

It follows from (3.4) and (3.8) that the Brun-type formalism of the first moment is given by

$$\begin{aligned}\left\langle (xy)_t \right\rangle = \frac{-1}{2\pi}\int_{-\pi}^{\pi}\int_{-\pi}^{\pi}\frac{dk_x dk_y}{2\pi}\sum_{j=1}^{t}\sum_{k=1}^{j-1} Tr\left\{Z\,L^{j-k}{}_{k_x,k_y}\left(ZL^{k}{}_{k_x,k_y}\rho_0\right)\right\} + Tr\left\{Z\,L^{j-k}{}_{k_x,k_y}\left(L^{k}{}_{k_x,k_y}\rho_0\right)Z\right\} + \\ \frac{-1}{2\pi}\int_{-\pi}^{\pi}\int_{-\pi}^{\pi}\frac{dk_x dk_y}{2\pi}\sum_{j=1}^{t} Tr\left(L^{j}{}_{k_x,k_y}\rho_0\right)\end{aligned} \tag{42}$$

## 4. Application of the Method of General Decoherence in Two Dimensions

In this section, as an application, we study the effects of coin-position decoherence on the quantum walk. In particular, we study the two-dimensional quantum walk in the presence of broken line noise, a case for which the coin noise and the position noise are inseparable.

The quantum walk in the presence of broken line noise have been studied by a number of authors, Romanelli et.al [25] calculated the diffusion co-efficient numerically in the case of the one-dimensional quantum walk, whilst Annabestani et.al [23] were able to confirm analytically the numerical experiments of Romanelli et.al [26]. The crossover in the diffusive behavior from the quantum walk to the classical walk is explained. In particular, in the presence of broken line noise diffusion in the quantum realm is much greater than diffusion in the classical realm. In this section we attempt to show how the broken line noise can lead to decoherence in the two-dimensional quantum walk, the work in the one-dimensional case implies the crossover in the diffusive behavior from the quantum realm to the classical realm depends on the frequency of noise. To that end to describe a genuine diffusive process we consider the classical version of the broken line model and give the diffusion co-efficient. In the quantum case we conjecture the diffusion co-efficient, and leave it as an exercise to verify or refute the proposal.

### 4.1. Basic Notions and Notations

In this section we introduce the notions and notations relative to studying the broken line model for the discrete-time quantum walk in the plane, some of which have been encountered earlier on. Recall that in the two dimensional quantum walk we have four degrees of freedom in the coin space which is spanned by $\{|L\rangle, |R\rangle, |U\rangle, |D\rangle\}$. The position space on the other hand is spanned by $\{|i\rangle, |j\rangle \;\; i, j = -\infty, \ldots \infty\}$. Again denoting the coin space by the Hilbert subspace $H_C$ and the position space by the Hilbert subspace $H_P$, the Hilbert space of the whole system is given by $H_P \otimes H_C$. In one step of the two-dimensional quantum walk we first make superposition on the coin space with coin operator $U_C$ and after that we move the particle according to the coin state with the translation operator $S$ as follows $U_w = S \cdot (I \otimes U_c)$, where

$$S = \sum_{x,y} \{|x-1, y\rangle\langle x, y| \otimes |L\rangle\langle L| + |x+1, y\rangle\langle x, y| \otimes |R\rangle\langle R| + |x, y-1\rangle\langle x, y| \otimes |D\rangle\langle D| + |x, y+1\rangle\langle x, y| \otimes |U\rangle\langle U|\}.$$

In our study of the broken line model we will set the coin operator as $U_C = H^{**}\left(\frac{1}{2}, \frac{1}{2}\right)$, where $H^{**}(p,q)$ is the Hadamard operation given in [27]. The evolution of the quantum walk is defined by $|\psi(t+1)\rangle = U_W |\psi(t)\rangle$. The wave vector is expressed as the spinor

$$\left|\psi(t)\right\rangle = \sum_{x,y=-\infty}^{\infty} \begin{bmatrix} L_{x.y}(t) \\ R_{x,y}(t) \\ D_{x,y}(t) \\ U_{x,y}(t) \end{bmatrix} \left|x,y\right\rangle$$, where the quibit $\begin{bmatrix} L & R & D & U \end{bmatrix}^T$ has the first component associated to the left chirality of the particle, the second component associated to the right chirality of the particle, and the third and fourth component associated to the down and up chirality of the particle respectively. The states $\left|x,y\right\rangle$ are the eigenstates of the position operator corresponding to the site $(x,y)$ in the plane. The evolution of the walk corresponding to $\left|\psi(t+1)\right\rangle = U_W \left|\psi(t)\right\rangle$ can be written as the map

$$L_{x,y}(t+1) = \frac{1}{2}\left\{L_{x+1,y}(t) + R_{x+1,y}(t) + D_{x+1,y}(t) + U_{x+1,y}(t)\right\}$$

$$R_{x,y}(t+1) = \frac{1}{2}\left\{L_{x-1,y}(t) - R_{x-1,y}(t) + D_{x-1,y}(t) - U_{x-1,y}(t)\right\}$$

$$D_{x,y}(t+1) = \frac{1}{2}\left\{L_{x,y+1}(t) + R_{x,y+1}(t) - D_{x,y+1}(t) - U_{x,y+1}(t)\right\}$$

$$U_{x,y}(t+1) = \frac{1}{2}\left\{L_{x,y-1}(t) - R_{x,y-1}(t) - D_{x,y-1}(t) + U_{x,y-1}(t)\right\}$$

The probability distribution of the walker position at time $t$ is given by

$$P_{x,y}(t) = \left|L_{x,y}(t)\right|^2 + \left|R_{x,y}(t)\right|^2 + \left|D_{x,y}(t)\right|^2 + \left|U_{x,y}(t)\right|^2$$

### 4.2. The Broken Line Model

Suppose at time $t$, a given site $(x,y)$ has one or more of the links connecting it to its neighboring sites broken. If site $(x,y)$ has no broken links, the evolution of the quantum walk progresses naturally. If one or more links at site $(x,y)$ are broken, then the evolution of the quantum walk must be modified accordingly. Note that site $(x,y)$ has four neighboring sites namely $(x-1,y)$, $(x+1,y)$, $(x,y-1)$, and $(x,y+1)$. Let us label the links connecting the neighboring sites as follows:

a) Let LINK I be the path from site $(x, y)$ to $(x+1, y)$
b) Let LINK II be the path from site $(x, y)$ to $(x-1, y)$
c) Let LINK III be the path from site $(x, y)$ to $(x, y+1)$
d) Let LINK IV be the path from site $(x, y)$ to $(x, y-1)$

Considering all possible combinations of the links, we see that there are fifteen different ways the evolution of the walk may proceed, if they are broken. In each case the corresponding transformation on the spinor components is as follows:

**Case 1:** If LINK I is broken

$$L_{x,y}(t+1)=\frac{1}{2}\left\{-L_{x,y}(t)+R_{x,y}(t)+D_{x,y}(t)+U_{x,y}(t)\right\}$$
$$R_{x,y}(t+1)=\frac{1}{2}\left\{L_{x-1,y}(t)-R_{x-1,y}(t)+D_{x-1,y}(t)-U_{x-1,y}(t)\right\}$$
$$D_{x,y}(t+1)=\frac{1}{2}\left\{L_{x,y+1}(t)+R_{x,y+1}(t)-D_{x,y+1}(t)-U_{x,y+1}(t)\right\}$$
$$U_{x,y}(t+1)=\frac{1}{2}\left\{L_{x,y-1}(t)-R_{x,y-1}(t)-D_{x,y-1}(t)+U_{x,y-1}(t)\right\}$$

**Case 2:** If LINK II is broken

$$L_{x,y}(t+1)=\frac{1}{2}\left\{L_{x+1,y}(t)+R_{x+1,y}(t)+D_{x+1,y}(t)+U_{x+1,y}(t)\right\}$$
$$R_{x,y}(t+1)=\frac{1}{2}\left\{L_{x,y}(t)+R_{x,y}(t)+D_{x,y}(t)-U_{x,y}(t)\right\}$$
$$D_{x,y}(t+1)=\frac{1}{2}\left\{L_{x,y+1}(t)+R_{x,y+1}(t)-D_{x,y+1}(t)-U_{x,y+1}(t)\right\}$$
$$U_{x,y}(t+1)=\frac{1}{2}\left\{L_{x,y-1}(t)-R_{x,y-1}(t)-D_{x,y-1}(t)+U_{x,y-1}(t)\right\}$$

**Case 3:** If LINK III is broken

$$L_{x,y}(t+1)=\frac{1}{2}\left\{L_{x+1,y}(t)+R_{x+1,y}(t)+D_{x+1,y}(t)+U_{x+1,y}(t)\right\}$$
$$R_{x,y}(t+1)=\frac{1}{2}\left\{L_{x-1,y}(t)-R_{x-1,y}(t)+D_{x-1,y}(t)-U_{x-1,y}(t)\right\}$$
$$D_{x,y}(t+1)=\frac{1}{2}\left\{L_{x,y}(t)+R_{x,y}(t)+D_{x,y}(t)-U_{x,y}(t)\right\}$$

$$U_{x,y}(t+1)=\frac{1}{2}\left\{L_{x,y-1}(t)-R_{x,y-1}(t)-D_{x,y-1}(t)+U_{x,y-1}(t)\right\}$$

**<u>Case 4:</u>** If LINK IV is broken

$$L_{x,y}(t+1)=\frac{1}{2}\left\{L_{x+1,y}(t)+R_{x+1,y}(t)+D_{x+1,y}(t)+U_{x+1,y}(t)\right\}$$
$$R_{x,y}(t+1)=\frac{1}{2}\left\{L_{x-1,y}(t)-R_{x-1,y}(t)+D_{x-1,y}(t)-U_{x-1,y}(t)\right\}$$
$$D_{x,y}(t+1)=\frac{1}{2}\left\{L_{x,y+1}(t)+R_{x,y+1}(t)-D_{x,y+1}(t)-U_{x,y+1}(t)\right\}$$
$$U_{x,y}(t+1)=\frac{1}{2}\left\{L_{x,y}(t)-R_{x,y}(t)-D_{x,y}(t)-U_{x,y}(t)\right\}$$

**<u>Case 5:</u>** If LINK I and LINK II are broken

$$L_{x,y}(t+1)=\frac{1}{2}\left\{-L_{x,y}(t)+R_{x,y}(t)+D_{x,y}(t)+U_{x,y}(t)\right\}$$
$$R_{x,y}(t+1)=\frac{1}{2}\left\{L_{x,y}(t)+R_{x,y}(t)+D_{x,y}(t)-U_{x,y}(t)\right\}$$
$$D_{x,y}(t+1)=\frac{1}{2}\left\{L_{x,y+1}(t)+R_{x,y+1}(t)-D_{x,y+1}(t)-U_{x,y+1}(t)\right\}$$
$$U_{x,y}(t+1)=\frac{1}{2}\left\{L_{x,y-1}(t)-R_{x,y-1}(t)-D_{x,y-1}(t)+U_{x,y-1}(t)\right\}$$

**<u>Case 6:</u>** If LINK I and LINK III are broken

$$L_{x,y}(t+1)=\frac{1}{2}\left\{-L_{x,y}(t)+R_{x,y}(t)+D_{x,y}(t)+U_{x,y}(t)\right\}$$
$$R_{x,y}(t+1)=\frac{1}{2}\left\{L_{x-1,y}(t)-R_{x-1,y}(t)+D_{x-1,y}(t)-U_{x-1,y}(t)\right\}$$
$$D_{x,y}(t+1)=\frac{1}{2}\left\{L_{x,y}(t)+R_{x,y}(t)+D_{x,y}(t)-U_{x,y}(t)\right\}$$
$$U_{x,y}(t+1)=\frac{1}{2}\left\{L_{x,y-1}(t)-R_{x,y-1}(t)-D_{x,y-1}(t)+U_{x,y-1}(t)\right\}$$

**<u>Case 7:</u>** If LINK I and IV are broken

$$L_{x,y}(t+1)=\frac{1}{2}\left\{-L_{x,y}(t)+R_{x,y}(t)+D_{x,y}(t)+U_{x,y}(t)\right\}$$
$$R_{x,y}(t+1)=\frac{1}{2}\left\{L_{x-1,y}(t)-R_{x-1,y}(t)+D_{x-1,y}(t)-U_{x-1,y}(t)\right\}$$
$$D_{x,y}(t+1)=\frac{1}{2}\left\{L_{x,y+1}(t)+R_{x,y+1}(t)-D_{x,y+1}(t)-U_{x,y+1}(t)\right\}$$
$$U_{x,y}(t+1)=\frac{1}{2}\left\{L_{x,y}(t)-R_{x,y}(t)-D_{x,y}(t)-U_{x,y}(t)\right\}$$

**<u>Case 8:</u>** If LINK  II and III are broken

$$L_{x,y}(t+1)=\frac{1}{2}\left\{L_{x+1,y}(t)+R_{x+1,y}(t)+D_{x+1,y}(t)+U_{x+1,y}(t)\right\}$$
$$R_{x,y}(t+1)=\frac{1}{2}\left\{L_{x,y}(t)+R_{x,y}(t)+D_{x,y}(t)-U_{x,y}(t)\right\}$$
$$D_{x,y}(t+1)=\frac{1}{2}\left\{L_{x,y}(t)+R_{x,y}(t)+D_{x,y}(t)-U_{x,y}(t)\right\}$$
$$U_{x,y}(t+1)=\frac{1}{2}\left\{L_{x,y-1}(t)-R_{x,y-1}(t)-D_{x,y-1}(t)+U_{x,y-1}(t)\right\}$$

**<u>Case 9:</u>** If LINK II and IV are broken

$$L_{x,y}(t+1)=\frac{1}{2}\left\{L_{x+1,y}(t)+R_{x+1,y}(t)+D_{x+1,y}(t)+U_{x+1,y}(t)\right\}$$
$$R_{x,y}(t+1)=\frac{1}{2}\left\{L_{x,y}(t)+R_{x,y}(t)+D_{x,y}(t)-U_{x,y}(t)\right\}$$
$$D_{x,y}(t+1)=\frac{1}{2}\left\{L_{x,y+1}(t)+R_{x,y+1}(t)-D_{x,y+1}(t)-U_{x,y+1}(t)\right\}$$
$$U_{x,y}(t+1)=\frac{1}{2}\left\{L_{x,y}(t)-R_{x,y}(t)-D_{x,y}(t)-U_{x,y}(t)\right\}$$

**<u>Case 10:</u>** If LINK III and IV are broken

$$L_{x,y}(t+1)=\frac{1}{2}\left\{L_{x+1,y}(t)+R_{x+1,y}(t)+D_{x+1,y}(t)+U_{x+1,y}(t)\right\}$$
$$R_{x,y}(t+1)=\frac{1}{2}\left\{L_{x-1,y}(t)-R_{x-1,y}(t)+D_{x-1,y}(t)-U_{x-1,y}(t)\right\}$$
$$D_{x,y}(t+1)=\frac{1}{2}\left\{L_{x,y}(t)+R_{x,y}(t)+D_{x,y}(t)-U_{x,y}(t)\right\}$$
$$U_{x,y}(t+1)=\frac{1}{2}\left\{L_{x,y}(t)-R_{x,y}(t)-D_{x,y}(t)-U_{x,y}(t)\right\}$$

**<u>Case 11:</u>** If LINK I, II, and III are broken

$$L_{x,y}(t+1)=\frac{1}{2}\left\{-L_{x,y}(t)+R_{x,y}(t)+D_{x,y}(t)+U_{x,y}(t)\right\}$$
$$R_{x,y}(t+1)=\frac{1}{2}\left\{L_{x,y}(t)+R_{x,y}(t)+D_{x,y}(t)-U_{x,y}(t)\right\}$$
$$D_{x,y}(t+1)=\frac{1}{2}\left\{L_{x,y}(t)+R_{x,y}(t)+D_{x,y}(t)-U_{x,y}(t)\right\}$$
$$U_{x,y}(t+1)=\frac{1}{2}\left\{L_{x,y-1}(t)-R_{x,y-1}(t)-D_{x,y-1}(t)+U_{x,y-1}(t)\right\}$$

**<u>Case 12:</u>** If LINK I, II, and IV are broken

$$L_{x,y}(t+1)=\frac{1}{2}\left\{-L_{x,y}(t)+R_{x,y}(t)+D_{x,y}(t)+U_{x,y}(t)\right\}$$
$$R_{x,y}(t+1)=\frac{1}{2}\left\{L_{x,y}(t)+R_{x,y}(t)+D_{x,y}(t)-U_{x,y}(t)\right\}$$
$$D_{x,y}(t+1)=\frac{1}{2}\left\{L_{x,y+1}(t)+R_{x,y+1}(t)-D_{x,y+1}(t)-U_{x,y+1}(t)\right\}$$
$$U_{x,y}(t+1)=\frac{1}{2}\left\{L_{x,y}(t)-R_{x,y}(t)-D_{x,y}(t)-U_{x,y}(t)\right\}$$

**<u>Case 13:</u>** If LINK I, III, IV are broken

$$L_{x,y}(t+1)=\frac{1}{2}\left\{-L_{x,y}(t)+R_{x,y}(t)+D_{x,y}(t)+U_{x,y}(t)\right\}$$
$$R_{x,y}(t+1)=\frac{1}{2}\left\{L_{x-1,y}(t)-R_{x-1,y}(t)+D_{x-1,y}(t)-U_{x-1,y}(t)\right\}$$
$$D_{x,y}(t+1)=\frac{1}{2}\left\{L_{x,y}(t)+R_{x,y}(t)+D_{x,y}(t)-U_{x,y}(t)\right\}$$
$$U_{x,y}(t+1)=\frac{1}{2}\left\{L_{x,y}(t)-R_{x,y}(t)-D_{x,y}(t)-U_{x,y}(t)\right\}$$

**<u>Case 14:</u>** If LINK II, III, and IV are broken

$$L_{x,y}(t+1)=\frac{1}{2}\left\{L_{x+1,y}(t)+R_{x+1,y}(t)+D_{x+1,y}(t)+U_{x+1,y}(t)\right\}$$
$$R_{x,y}(t+1)=\frac{1}{2}\left\{L_{x,y}(t)+R_{x,y}(t)+D_{x,y}(t)-U_{x,y}(t)\right\}$$
$$D_{x,y}(t+1)=\frac{1}{2}\left\{L_{x,y}(t)+R_{x,y}(t)+D_{x,y}(t)-U_{x,y}(t)\right\}$$
$$U_{x,y}(t+1)=\frac{1}{2}\left\{L_{x,y}(t)-R_{x,y}(t)-D_{x,y}(t)-U_{x,y}(t)\right\}$$

**<u>Case 15:</u>** If LINK I, II, III, IV are broken

$$L_{x,y}(t+1) = \frac{1}{2}\left\{ -L_{x,y}(t) + R_{x,y}(t) + D_{x,y}(t) + U_{x,y}(t) \right\}$$
$$R_{x,y}(t+1) = \frac{1}{2}\left\{ L_{x,y}(t) + R_{x,y}(t) + D_{x,y}(t) - U_{x,y}(t) \right\}$$

$$D_{x,y}(t+1) = \frac{1}{2}\left\{ L_{x,y}(t) + R_{x,y}(t) + D_{x,y}(t) - U_{x,y}(t) \right\}$$
$$U_{x,y}(t+1) = \frac{1}{2}\left\{ L_{x,y}(t) - R_{x,y}(t) - D_{x,y}(t) - U_{x,y}(t) \right\}$$

Let us remark that in these cases the evolution of the quantum walk consists of a sequence of unitary transformations applied to the initial wave function $\left|\Psi(t)\right\rangle = U_t\, U_{t-1} \ldots U_1 \left|\Psi(0)\right\rangle$ and the form of each $U_k$ depends on the topology of the plane at the particular time step [24].

It also follows that the set of equations in the fifteen cases above and those in the case of no broken lines preserve the norm of the wave function, $\sum P_{x,y}(t) = 1$.

### 4.3. Emergence of Classical Diffusion

To give the classical evolution of the broken line model, we need to express the evolution equations in each of the fifteen cases in the previous section, including the case where there are no broken links in terms of the probability distribution for the walker at time $t$, which we defined earlier on as

$$P_{x,y}(t) = \left|L_{x,y}(t)\right|^2 + \left|R_{x,y}(t)\right|^2 + \left|D_{x,y}(t)\right|^2 + \left|U_{x,y}(t)\right|^2 \tag{43}$$

From equation (43) the evolution equations in the case of no broken links can be written in terms of the probability distribution of the walker as

$$P_{x,y}(t+1) = \frac{1}{4}\left[ P_{x+1,y}(t) + P_{x-1,y}(t) + P_{x,y+1}(t) + P_{x,y-1}(t) \right] + other\ terms \tag{44}$$

In the case of one or more broken links, the evolution equations in each of the fifteen cases in the previous section can be written in terms of the probability distribution of the walker, as follows

$$P_{x,y}(t+1)=\frac{1}{4}\left[P_{x,y}(t)+P_{x-1,y}(t)+P_{x,y+1}(t)+P_{x,y-1}(t)\right]+other\ terms \tag{44}$$

$$P_{x,y}(t+1)=\frac{1}{4}\left[P_{x+1,y}(t)+P_{x,y}(t)+P_{x,y+1}(t)+P_{x,y-1}(t)\right]+other\ terms \tag{45}$$

$$P_{x,y}(t+1)=\frac{1}{4}\left[P_{x+1,y}(t)+P_{x-1,y}(t)+P_{x,y}(t)+P_{x,y-1}(t)\right]+other\ terms \tag{46}$$

$$P_{x,y}(t+1)=\frac{1}{4}\left[P_{x+1,y}(t)+P_{x-1,y}(t)+P_{x,y+1}(t)+P_{x,y}(t)\right]+other\ terms \tag{47}$$

$$P_{x,y}(t+1)=\frac{1}{4}\left[P_{x,y}(t)+P_{x,y}(t)+P_{x,y+1}(t)+P_{x,y-1}(t)\right]+other\ terms \tag{48}$$

$$P_{x,y}(t+1)=\frac{1}{4}\left[P_{x,y}(t)+P_{x-1,y}(t)+P_{x,y}(t)+P_{x,y-1}(t)\right]+other\ terms \tag{49}$$

$$P_{x,y}(t+1)=\frac{1}{4}\left[P_{x,y}(t)+P_{x-1,y}(t)+P_{x,y+1}(t)+P_{x,y}(t)\right]+other\ terms \tag{50}$$

$$P_{x,y}(t+1)=\frac{1}{4}\left[P_{x+1,y}(t)+P_{x,y}(t)+P_{x,y}(t)+P_{x,y-1}(t)\right]+other\ terms \tag{51}$$

$$P_{x,y}(t+1)=\frac{1}{4}\left[P_{x+1,y}(t)+P_{x,y}(t)+P_{x,y+1}(t)+P_{x,y}(t)\right]+other\ terms \tag{52}$$

$$P_{x,y}(t+1)=\frac{1}{4}\left[P_{x+1,y}(t)+P_{x-1,y}(t)+P_{x,y}(t)+P_{x,y}(t)\right]+other\ terms \tag{53}$$

$$P_{x,y}(t+1)=\frac{1}{4}\left[P_{x,y}(t)+P_{x,y}(t)+P_{x,y}(t)+P_{x,y-1}(t)\right]+other\ terms \tag{54}$$

$$P_{x,y}(t+1)=\frac{1}{4}\left[P_{x,y}(t)+P_{x,y}(t)+P_{x,y+1}(t)+P_{x,y}(t)\right]+other\ terms \tag{55}$$

$$P_{x,y}(t+1)=\frac{1}{4}\left[P_{x,y}(t)+P_{x-1,y}(t)+P_{x,y}(t)+P_{x,y}(t)\right]+other\ terms \tag{56}$$

$$P_{x,y}(t+1)=\frac{1}{4}\left[P_{x+1,y}(t)+P_{x,y}(t)+P_{x,y}(t)+P_{x,y}(t)\right]+other\ terms \tag{57}$$

$$P_{x,y}(t+1) = P_{x,y}(t) + other\ terms \tag{58}$$

Note that in equations (44)-(58) the "other terms" take into account the quantum coherence effects and are responsible for the essential differences between the classical and quantum random walks. Under the assumption of a complete decay of correlations their contribution is neglible and the classical description emerges [28, 29].

In order to see the classical evolution, we need to combine equations (44)-(58) into one single evolution equation with the appropriate statistical weights. Let $p$ be the probability that a given site has an adjacent link broken, then the probability that a given site has no adjacent broken links is given by $\Pi_0 = (1-p)^4$, that it has only one adjacent broken link is given by $\Pi_1 = p^3(1-p)$, that it has two adjacent broken links is given by $\Pi_2 = p^2(1-p)^2$, that is has three adjacent broken links is given by $\Pi_3 = p(1-p)^3$, and that it is isolated, meaning all the adjacent links are broken is $\Pi_4 = p^4$, note that $\Pi_0 + 4\Pi_1 + 6\Pi_2 + 5\Pi_3 + \Pi_4 = 1$. The resulting classical evolution equation for $P_{x,y}(t)$ is given by $P_{x,y}(t+1) = pP_{x,y}(t) + \frac{1}{4}(1-p)\left[P_{x-1,y}(t) + P_{x+1,y}(t) + P_{x,y-1}(t) + P_{x,y+1}(t)\right]$ where the diffusion coefficient is given by $D_{classical} = \frac{1}{4}(1-p)$, this can also been seen by changing to continuous time and position variables $(x, y, t)$, in terms of which $P_{x,y}(t+1) = pP_{x,y}(t) + \frac{1}{4}(1-p)\left[P_{x-1,y}(t) + P_{x+1,y}(t) + P_{x,y-1}(t) + P_{x,y+1}(t)\right]$ becomes the diffusion equation in two dimensions, namely, $\frac{\partial P}{\partial t} = D_{classical}\left(\frac{\partial^2 P}{\partial x^2} + \frac{\partial^2 P}{\partial y^2}\right)$. Thus the classical version of the broken link model with complete correlation decay which starts at the unbiased random walker value $\frac{1}{4}$ for $p = 0$ decreases as the frequency of the broken links increases. Based on the value of the classical diffusion coefficient for the broken link model in two dimensions, and the results for the diffusion coefficient in the broken link model in one dimension for both the classical and quantum case, we speculate the following.

**Conjecture 1:** The diffusion coefficient in the quantum realm is proportional to $\frac{1-p}{p^2}$ , where the constant of proportionality may or may not be a function of $p$

**Conjecture 2:** If Conjecture 1 is true, then our result for the diffusion coefficient in the classical case implies that the diffusion coefficient of the decoherent quantum walk is always larger than the diffusion coefficient in the classical case.

The proof of the above conjectures is left to the reader. However, we note that the proof of the first conjecture can be carried out experimentally by mimicking the work of Romanelli et.al [25] . The second conjecture can be thought of more or less a corollary of the first conjecture.

## 5. Concluding Remarks

In this paper we have studied general decoherence in quantum walks in two dimensions. We have obtained analytic expressions for the first and second moments in the presence of decoherence. We have shown in the coin decoherence, that the general two dimensional formalism is of the form considered by Brun et.al [30] via the method of Annabestani et.al [23]. We have also given an application of the method of general decoherence in two dimensions by considering the broken line decoherence, and obtained the diffusion coefficient in the classical case. In the quantum case we have conjectured the diffusion co-efficient which if holds true confirms the crossover in the diffusive behavior with respect to the quantum and classical realms as seen in one-dimension.